\documentclass[aps,prl,twocolumn,showpacs,superscriptaddress]{revtex4}
\begin{document}
\title{Information Entropy in Cosmology\footnote{a short version of this 
{\em Letter} received honorable mention in the GRG essay competition 2003}}
\author{Akio Hosoya\footnote{ahosoya@th.phys.titech.ac.jp}}
\affiliation{Department of Physics,
Tokyo Institute of Technology, Oh--Okayama, Meguro--ku, Tokyo 152--0033, Japan}
\author{Thomas Buchert\footnote{buchert@theorie.physik.uni-muenchen.de}}
\affiliation{Theoretische Physik, Ludwig--Maximilians--Universit\"at,
Theresienstr. 37, D--80333 M\"unchen, Germany}
\affiliation{Department of Physics,
Tokyo Institute of Technology, Oh--Okayama, Meguro--ku, Tokyo 152--0033, Japan}
\affiliation{Department of Physics and Research Center for the Early Universe
(RESCEU), School of Science, The University of Tokyo, Tokyo 113--0033, Japan}
\author{Masaaki Morita\footnote{masaaki@cosmos.phys.ocha.ac.jp}}
\affiliation{Department of Physics, Ochanomizu University, Ohtsuka, 
Bunkyo--ku, Tokyo 112--8610, Japan}

\affiliation{Advanced Research Institute for Science and Engineering, 
Waseda University, Ohkubo, Shinjuku--ku, Tokyo 169--8555, Japan}

\bigskip\bigskip

\begin{abstract}
The effective evolution of an inhomogeneous cosmological model may be described 
in terms of spatially averaged variables. We point out that in this context, 
quite naturally, a measure arises which is identical to a fluid model of the 
`Kullback--Leibler Relative Information Entropy', expressing the distinguishability 
of the local inhomogeneous mass density field from its spatial average on arbitrary 
compact domains. We discuss the time--evolution of `effective information' and 
explore some implications. We conjecture that the information content of the 
Universe -- measured by Relative Information Entropy of a cosmological model 
containing dust matter -- is increasing.
\end{abstract}
\maketitle
\section{A Measure of Inhomogeneity in the Universe}
Cosmology is based on the hypothesis of simplicity called the 
cosmological principle, i.e. homogeneity and isotropy. 
The departure of the actual mass distribution from the 
homogeneous universe model is quantified in terms of density contrast or a
statistical quantity like the two--point correlation function, which both have been
studied either by perturbation theory or numerical simulations. Behind these 
investigations there is a belief that the Universe is homogeneous on some large 
enough scale. This belief has to be quantitatively confronted with  
observation, explicitly introducing a measure of inhomogeneity for a domain 
of the Universe.

\smallskip

In this {\em Letter} we propose a measure which quantifies the distinguishability  of 
the actual mass distribution from its spatial average, borrowing  a well--known concept 
in standard information theory. Suppose we are told that the probability
distribution is $\{q_{i}\}$ and would like to examine how close
this distribution is to the actual one $\{p_{i}\}$ by carrying out observations
or coin tossing; the relevant quantity in information theory is the 
{\it relative entropy},
\begin{equation}
{\cal S} \lbrace p || q\rbrace = \sum_{i} p_{i}\ln \frac{p_{i}}{q_{i}}\;\;,
\end{equation}
which is positive for ${q_{i}}\neq {p_{i}}$, and zero if the
actual distribution $\{p_{i}\}$ agrees with the presumed one $\{q_{i}\}$.
Note that this relative entropy is not symmetric for the two distributions  
$\{p_{i}\}$ and $\{q_{i}\}$.
It is known that this measure always decreases or stays the same under 
Markovian stochastic processes (i.e., a {\it linear} positive map). 
Namely, the actual distribution becomes less and less 
distinguishable from the priorly informed distribution due to the random process. 
In cosmology we are interested in how the real matter distribution is
different from its spatial average. For a continuum the relevant quantity would be
\begin{equation}
\label{entropy}
\frac{{\cal S} \lbrace \varrho || \langle\varrho\rangle_{\cal D}\rbrace}{V_{\cal D}} \;=\; 
\Bigl\langle \varrho \ln \frac{\varrho}{\langle\varrho\rangle_{\cal D}}\Bigr\rangle_{\cal D}\;\;,
\end{equation}
where $\varrho$ is the actual distribution and $\langle \cdots \rangle_{\cal D}$
its spatial average in the volume $V_{\cal D}$ on the compact domain $\cal D$
of the manifold $\Sigma$.
We shall conjecture that 
the measure ${\cal S} \lbrace \varrho || \langle\varrho\rangle_{\Sigma}\rbrace$
continues to grow indefinitely, if $\Sigma$ is compact. 

The resolution of the apparent discrepancy 
between the gravitational system and the ordinary stochastic system will be, 
(i) we are considering in cosmology a non--isolated system defined by a comoving 
region $\cal D$ in contrast 
to an isolated system for an ordinary stochastic process, and (ii) the time evolution
dictated by Einstein's equations induces a negative feed--back due to the attractive
nature of the gravitational force, which tends to make the matter distribution 
more and more inhomogeneous.


\section{Deduction of the Measure}

To begin with let us emphasize that the functional (\ref{entropy}), known as the 
`Kullback--Leibler Relative Information Entropy' 
({\it cf} \cite{kullback}, \cite{kullback_leibler}, \cite{cover:entropy}) 
is not assumed as a measure {\em a priori}, rather it can be {\em deduced} 
from a fundamental kinematical relation that refers to the {\em non--commutativity} 
of two operations: spatially averaging and evolving the material mass density field.
The specific form of the information measure is, thus, 
inherently determined by the physical problem at hand, and does not need to 
be justified empirically or axiomatically as is the common status of 
information measures in the literature.

We define the averaging operation in terms of Riemannian volume
integration, restricting attention to scalar functions $\Psi (t,X^i)$, 
\begin{equation}
\label{average}
\langle \Psi (t, X^i)\rangle_{\cal D}: = 
\frac{1}{V_{\cal D}}\int_{\cal D} \sqrt{g} d^3 X \;\;\Psi (t, X^i) \;\;\;,
\end{equation}
with the Riemannian volume element $d\mu_g := \sqrt{g} d^3 X$, $g:=\det(g_{ij})$, and 
the volume of an arbitrary compact domain, $V_{\cal D}(t) : = \int_{\cal D} \sqrt{g} d^3 X$;
$X^i$ are coordinates in a $t=const.$ hypersurface (with $3-$metric $g_{ij}$) that are
comoving with fluid elements of dust:
\begin{equation}
ds^2 = -dt^2 + g_{ij}dX^idX^j \;\;.
\end{equation}
It is evident from the above setting that we predefine a simple time--orthogonal foliation
(which restricts the matter to an irrotational dust continuum)
in order to simplify the framework in which we discuss our measure as a concept of a 
{\em spatial} average. We wish to emphasize that the formalism below could be carried
over to more general settings (e.g. to perfect fluids or scalar fields ({\it
cf.} \cite{buchert:grgfluid}) with possibly further interesting implications. 

The above--mentioned `non--commutativity' has been fruitfully exploited 
in previous work on the averaging problem of inhomogeneous cosmologies
\cite{buchert:average, buchert:grgdust, buchert:grgfluid, buchert:onaverage,
buchertcarforaPRL},
and can be compactly written in terms of a {\em commutation rule} for the averaging of a
scalar field $\Psi$:
\begin{eqnarray}
\label{commutationrule}
\langle \Psi{\dot \rangle}_{\cal D} - \langle{\dot \Psi}\rangle_{\cal D}
= \langle \Psi\theta\rangle_{\cal D} - 
\langle \Psi\rangle_{\cal D}\langle\theta\rangle_{\cal D} \nonumber \\
=\langle\Psi\delta\theta\rangle_{\cal D}
=\langle\theta\delta\Psi\rangle_{\cal D}
=\langle\delta\Psi\delta\theta\rangle_{\cal D}\;\;\;,
\end{eqnarray}
where $\theta$ denotes the local expansion rate (as minus the trace of the extrinsic 
curvature of the hypersurfaces $t=const.$).
We have rewritten the r.h.s. of the first equality in terms of the deviations of the 
local fields from their spatial averages, $\delta\Psi := \Psi - \langle\Psi\rangle_{\cal D}$ and 
$\delta\theta := \theta - \langle\theta\rangle_{\cal D}$.

\bigskip

The key--statement of the {\em commutation rule} (\ref{commutationrule}) is that the operations 
{\it spatial averaging} and {\it time evolution} do not commute. 
In cosmology we may think of initial conditions at the epoch of last scattering, when the 
fluctuations imprinted on the Cosmic Microwave Background are considered to be 
averaged--out on a restframe of a standard Friedmann--Lema\^\i tre--Robertson--Walker (FLRW)
cosmology. In this picture the evolution of the Universe is described by first averaging--out
(or ignoring) inhomogeneities and then evolving the average distribution 
by a homogeneous (in the above case homogeneous--isotropic) universe model. A realistic model
would first evolve the inhomogeneous fields and, at the present epoch, the resulting fields
would have to be evaluated by spatial averaging to obtain the final values of, e.g., the averaged
density field. In particular, this comment applies to all cosmological parameters (see, e.g.,
\cite{buchert:grgdust} and \cite{buchertcarforaPRL}).
Let us illustrate this statement for the mass density field. 
Setting $\Psi = \varrho$,
Eq.~(\ref{commutationrule}) reads:
\begin{equation}
\label{commutationdensity1}
\langle \varrho{\dot\rangle}_{\cal D} +
\langle\theta\rangle_{\cal D}\langle \varrho\rangle_{\cal D} = 
\langle{\dot \varrho} + \theta\varrho\rangle_{\cal D}  
\;\;\;.
\end{equation}
Since the r.h.s. vanishes due to the continuity equation, we also have a 
continuity equation for the averages:
\begin{equation}
\label{continuity2}
\langle \varrho{\dot\rangle}_{\cal D} +
\langle\theta\rangle_{\cal D}\langle \varrho\rangle_{\cal D} = 0\;\;\;,
\end{equation}
which simply expresses the conservation of the total material mass, 
$M_{\cal D} = \int_{\cal D} \sqrt{g} d^3 X\;\varrho$, in our comoving and
synchronous gauge setting.

A fairly general insight that, in principle, will not depend on some specialized setting, can be
obtained by rewriting Eq.~(\ref{commutationdensity1}):
the notion of `non--commutativity' mentioned above comes into the
fore by observing  that the time--evolution of the average density
does not coincide with the average of the locally evolved density:
\begin{equation}
\label{commutationdensity2}
\langle \varrho{\dot\rangle}_{\cal D} - \langle{\dot \varrho}\rangle_{\cal D}
= \langle \varrho\theta\rangle_{\cal D} - 
\langle \varrho\rangle_{\cal D}\langle\theta\rangle_{\cal D}
=\langle\delta\varrho\delta\theta\rangle_{\cal D}\;\;\;.
\end{equation}
For the fluctuation terms on the r.h.s., which would vanish 
in the FLRW model without any perturbation, 
we can give a deeper interpretation. For this end
let us ask, which functional will reproduce these terms
upon performing the time--derivative.
First, note that for the averaged expansion rate 
$\langle\theta\rangle_{\cal D}$ the corresponding functional
is the volume according to 
\begin{equation}
\label{averageexpansion}
\langle\theta\rangle_{\cal D} = \frac{{\dot V}_{\cal D}}{V_{\cal D}} =:3H_{\cal D} \;\;\;.
\end{equation} 
The latter equality demonstrates that this quantity may be regarded as an {\em effective 
Hubble function}, which will show up in our discussion later.

Interestingly, the answer is provided, for $\varrho > 0$, by the functional
 ${\cal S}\lbrace\varrho || \langle\varrho\rangle_{\cal D}\rbrace$, 
Eq.~(\ref{entropy}),
so that the source of non--commutativity in Eq.~(\ref{commutationdensity2}) 
is given (up to the sign) by the production
of {\it Relative Information Entropy}, defined as to measure the 
deviations from the average mass density due to the development of inhomogeneities:
\begin{equation}
\label{relativeentropy}
\langle \varrho{\dot\rangle}_{\cal D} - \langle {\dot\varrho}\rangle_{\cal D} =
 -\frac{{\dot {\cal S}\lbrace\varrho || \langle\varrho\rangle_{\cal
D}\rbrace}}{V_{\cal D}}\;\;.
\end{equation}
This measure can actually be inferred from its definition in phase space in terms of 
the one--particle distribution function for dust matter, i.e. 
the matter density multiplied by a delta--function distribution in 
velocity space \cite{hosoya:infoentropy}.
It is here, where generalizations of the matter model, e.g. supported by 
pressure, vorticity and/or velocity dispersion could be implemented, 
resulting in more general entropies after 
taking velocity moments in phase space. 
%

The reader may ask, whether this measure is superior to the density fluctuation
measure, which also provides a generally 
growing and positive--definite valuation of the density distribution. 
Let us give some answers to this question before we proceed.

A standard index of inhomogeneity in cosmology is the density contrast 
$\delta:=\frac{\delta\varrho}{\langle\varrho\rangle_{\cal D}}$ and the derived 
positive measure $(\Delta\varrho)^2 : = \langle \varrho^2 \rangle_{\cal D} - 
\langle\varrho\rangle_{\cal D}^2$. 
The Relative Information Entropy or
the distinguishability, Eq.~(\ref{entropy}), may have further implications
by exploiting results from information theory. At the present stage we do not 
claim that this measure is superior to the density fluctuation, but
rather it is {\it complementary}. This can be illustrated 
by pointing out that both measures are 
``cousins'' in a $1$--parameter family of inhomogeneity measures defined by

\begin{equation}
{\cal F}_{\alpha} \lbrace\varrho || \langle\varrho\rangle_{\cal D}\rbrace
:=\frac{\langle\varrho\rangle_{\cal D}}{\alpha}\left[\Bigl\langle
\left(\frac{\varrho}{\langle\varrho\rangle_{\cal D}}
\right)^{\alpha+1}\Bigr\rangle_{\cal D}-1\right]\;\;,\nonumber
\end{equation}
with $\alpha$ being a real parameter.
In the limit $\alpha\rightarrow 0$ the formula reproduces the relative entropy,
${\cal F}_{\alpha \rightarrow 0}\rightarrow {\cal S}/V_{\cal D}$,
whereas $\alpha=1$ reproduces the density fluctuation,
${\cal F}_{\alpha =1} = (\Delta\varrho)^2 / \langle\varrho\rangle_{\cal D}$.
This interpolating formula is known as the {\it Tsallis relative entropy}.
It should be emphasized that the limit $\alpha \rightarrow 0$ is singled out
as the only measure that exactly provides the source of non--commutativity
with regard to the density evolution. 

\section{Properties of the Measure}

The measure ${\cal S}\lbrace\varrho || \langle\varrho\rangle_{\cal D}\rbrace$
forms one of the central concepts in information theory \cite{cover:entropy};
${\cal S}=0$ (``zero structure'') is attained by the homogeneous mass 
distribution, $\varrho = \langle\varrho\rangle_{\cal D}$.

First, for strictly positive mass density, $\varrho > 0$, 
${\cal S}\lbrace\varrho || \langle\varrho\rangle_{\cal D}\rbrace$
is positive definite, which can be readily confirmed,
i.e. it is indeed a {\em measure}.


Let us have a closer look at the total time--derivative of our measure.
Following from what has been said above, we may  
write the total {\em Relative Information Entropy production} as follows:
\begin{equation}
\label{entropyproduction}
\frac{{\dot {\cal S}\lbrace\varrho || \langle\varrho\rangle_{\cal D}\rbrace}}{V_{\cal D}} 
= -\langle\delta\varrho\theta\rangle_{\cal D} = -\langle\varrho\delta\theta\rangle_{\cal D} 
= -\langle\delta\varrho\delta\theta\rangle_{\cal D} \;\;.
\end{equation}
The last quantity is bounded according to Schwarz' inequality, so that we obtain: 
\begin{equation}
\label{bound1}
\Big\vert\frac{{\dot {\cal S}\lbrace\varrho || \langle\varrho\rangle_{\cal D}\rbrace}}{V_{\cal D}} 
\Big\vert=|\langle\delta\varrho\delta\theta\rangle_{\cal D}| 
\;\le\; \Delta\varrho \Delta\theta \;\;,
\end{equation}
with the positive--definite fluctuation amplitudes
\begin{equation}
\Delta\varrho := \sqrt{\langle(\delta\varrho)^2\rangle}\;\;;\;\; \Delta\theta :=
\sqrt{\langle(\delta\theta)^2\rangle}\;\;.
\end{equation}
This inequality states that the temporal change of the ratio between the 
{\it distinguishability} of the density distribution from 
the homogeneous distribution and the volume is bounded by the density and 
expansion fluctuation amplitudes. We may say that the production of information in the 
Universe and its volume expansion are competing. 

\smallskip

We may look more closely at bounds as well as
kinematical and dynamical conditions for the total second
time--derivative of ${\cal S}\lbrace\varrho || \langle\varrho\rangle_{\cal D}\rbrace$. 
In \cite{hosoya:infoentropy} we give sufficient conditions for the {\em time--convexity}
of our measure. Let us put one of them into perspective.
We consider the question under which condition the time--derivative of the
Relative Information Entropy production is positive. A straightforward calculation 
provides:
\begin{equation}
\label{Pdot2}
\frac{{\ddot{\cal S}}}{V_{\cal D}} = 
-\langle\delta\varrho\delta{\dot\theta}\rangle_{\cal D} + \langle \varrho\rangle_{\cal D} 
(\Delta\theta)^2 \;\;.
\end{equation} 
Raychaudhuri's equation, 
\begin{equation}
\label{raychaudhuri}
\dot\theta = \Lambda - 4\pi G\varrho - \frac{1}{3}\theta^2 - 2\sigma^2\;\;,
\end{equation}
with the rate of shear $\sigma := \sqrt{\frac{1}{2}\sigma^i_{\;j} \sigma^j_{\;i}}$, the 
shear tensor $\sigma_{ij}$ being minus the trace--free part of the extrinsic curvature), 
together with the {\em commutation rule} (\ref{commutationrule}) yields:
\begin{eqnarray*}
\frac{{\ddot{\cal S}}}{V_{\cal D}}& = 4\pi G(\Delta 
\varrho)^2+\langle\varrho\rangle_{\cal D}(\Delta\theta)^2 + \frac{1} 
{3}\langle\delta\varrho\delta\theta^2\rangle_{\cal D}+2\langle\delta
\varrho\delta\sigma^2\rangle_{\cal D}\\
\geq & 4\pi G (\Delta \varrho)^2 - \Delta \varrho[\frac{1}{3} 
\Delta\theta^2 + 2\Delta\sigma^2] + \langle\varrho\rangle_{\cal D}
(\Delta\theta)^2\;\;.
\end{eqnarray*}
The r.h.s. is positive, if
\begin{equation}
\label{sufficient}
\frac{1}{2} \frac{\Delta(\frac{1}{3}\theta^2)+\Delta(2\sigma^2)} 
{\Delta\theta} \leq \sqrt{4\pi G\langle\varrho\rangle_{\cal D}} = \frac{1}{t_{F_{\cal D}}}\;\;,
\end{equation}
where $t_{F_{\cal D}}$ denotes the {\em effective free--fall time} on $\cal D$. 

Eq.~(\ref{sufficient}) provides a sufficient condition for the time--convexity of the Relative 
Information Entropy, which can be  met, if gravity dominates over expansion 
and shear fluctuations. Time--convexity implies that entropy 
production eventually becomes positive, i.e. the structure eventually 
surfaces and its rate of formation increases.

\section{Discussion and Conjecture}

Looking at Eq.~(\ref{entropyproduction}) we appreciate that
the source, i.e., the averaged Relative Information Entropy production
density, can be positive or negative. 
In cosmology, the processes of a relative accumulation of matter (cluster
formation) and a relative dilution of matter (void formation) create
{\it structure} compared with the average distribution. 
Following from Eq.~(\ref{entropyproduction}), 
information entropy is produced if, on average, there are overdense fluid elements
($\delta\varrho > 0$) which are contracting ($\theta < 0$), or underdense elements  
($\delta\varrho < 0$) which are expanding ($\theta > 0$), respectively.
With regard to cosmological structure formation these two states are generically 
encountered in a self--gravitating system, i.e., for large enough times and looking at 
some regional scale, an asymmetry of states is created due to the coupling of the expansion
rate to the rate of change of the density through the continuity equation.
We know from a calculation of the measure in linear perturbation theory
that the growing--mode solution supports  
states with $\lbrace \delta\varrho > 0\;,\;\theta < 0 \rbrace$ 
(contracting clusters) and $\lbrace \delta\varrho < 0\;,\;\theta > 0 \rbrace$ 
(expanding voids).
Thus, {\em for sufficiently large times}, i.e. when the decaying mode 
disappears, our measure will increase.

Looking at Eq.~(\ref{sufficient}) we conclude that also in the case of the 
second time--derivative we have the possibility of time--concavity of the
Relative Information Entropy.
However, we have evidence that, at least for large enough times and on sufficiently large
scales, time--convexity {\em always} holds for a self--gravitating continuum of dust. 
In particular, in the linear perturbation theory, our measure is {\em always} 
time--convex \cite{hosoya:infoentropy}.

We can illustrate roughly the physical content of the sufficient condition (\ref{sufficient})
as follows. Concentrating on the linear regime
by considering the case in which fluctuations of a quantity are small compared with their
average values, we may expand the quadratic expressions and keep only the leading terms:
\begin{equation}
\Delta(\theta^2) \;\approx \; 2|\langle \theta \rangle_{\cal D}|\Delta\theta \;\;\;;\;\;\;
\Delta(\sigma^2) \;\approx \; 2|\langle \sigma \rangle_{\cal D}|\Delta\sigma \;\;\;.
\end{equation}   
In this limit, if we additionally think of a large domain featuring approximately vanishing 
average shear, $\langle \sigma \rangle_{\cal D}\approx 0$, the sufficient condition 
(\ref{sufficient}) reduces to the inequality
\begin{equation}
\label{times}
t_{F_{\cal D}} \; \le \; t_{H_{\cal D}} =: | H_{\cal D}^{-1} | \;\;,
\end{equation}
i.e., if the {\em effective free--fall time} on $\cal D$ is smaller than the 
{\em effective Hubble time} $t_{H_{\cal D}}$, 
with $H_{\cal D}$ defined in Eq.~(\ref{averageexpansion}), then
time--convexity of our measure is ensured under the given assumptions.
The expectation that both positivity of the Relative Information Entropy
production and time--convexity, which are supported by the linear
perturbation results, will hold in the dust continuum {\em generically},
at least for large enough times and on sufficiently large scales of averaging,
establishes the following.

\bigskip
\noindent
{\em Conjecture:} 
The Relative Information Entropy of a dust matter model
${\cal S} \lbrace \varrho || \langle\varrho\rangle_{\Sigma}\rbrace$
is, for sufficiently large times, globally (i.e. averaged over the
whole compact manifold $\Sigma$) an increasing function of time.

\bigskip

We are currently
investigating nonlinear exact solutions for spherically--symmetric domains 
\cite{hosoya:infoentropy},
which may provide further support for our conjecture.

A note is in order as for the relation to observational constraints. In our context
a generalized form of Friedmann's differential equation governs the averaged expansion
(\ref{averageexpansion}), and a set of four effective cosmological parameters can be defined
\cite{buchert:grgdust}, \cite{buchertcarforaPRL}.
Assuming that, on sufficiently large scales of averaging, kinematical fluctuations 
and the averaged $3-$Ricci curvature have negligible contributions, 
then the sum of the cosmological parameters for the matter content
and the cosmological term have to add up to $1$; the former is indeed given by the 
fraction of the two competing times:
\begin{equation}
\Omega^m_{\cal D}:= \frac{8\pi G \langle \varrho\rangle_{\cal D}}{3 H_{\cal D}^2} = 
\frac{2}{3}\frac{t_{H_{\cal D}}^2}{t_{F_{\cal D}}^2} \;\;.
\end{equation}
Refering to observational results, e.g. by WMAP ({\em Wilkinson Microwave Anisotropy
Probe})\cite{WMAP}, its contribution is $\Omega^m_{\cal D}\approx 0.3$ and, thus, 
$t_{F_{\cal D}}$ is slightly larger than $t_{H_{\cal D}}$.
Note that this does not immediately imply that our measure is not
time--convex, because the condition (\ref{times}) derives from the sufficient 
condition (\ref{sufficient}),
which only provides a rough estimation and is not very stringent. 
On cosmological scales both times are indeed very similar, so that we 
should make the estimation tighter to see whether or not time--convexity 
holds; this we postpone to the future work \cite{hosoya:infoentropy}.

\smallskip

We contemplate that the measure that we propose in the present {\em Letter} not only 
incorporates an assessment of structure, but 
may turn out to be a fundamental quantity in many other respects, e.g. 
for the study of Black Holes and the Early Universe. 

\medskip

\begin{acknowledgments}
We would like to thank Alvaro Dom\'\i nguez, Jos\'e Gaite and Atsushi Taruya
for constant interest and discussions.
TB acknowledges hospitality during a visit in 2001 at the Hosoya Laboratory 
of the Tokyo Institute of Technology, where the main work on this subject
was done with support by The Japanese Society for the Promotion of 
Science (JSPS). He also acknowledges hospitality at the University of Tokyo with
support by the Research Center for the Early Universe (RESCEU, Tokyo),
COE Monkasho Grant, where this Letter was written. 
This work is also partially supported by the Sonderforschungsbereich SFB 375 
`Astroparticle physics' by the German science foundation DFG.
\end{acknowledgments}


\begin{thebibliography}{2004}

\bibitem{buchert:grgdust}
T. Buchert, G.\ R.\ G.\ {\bf 32}, 105 (2000).

\bibitem{buchert:grgfluid}
T. Buchert, G.\ R.\ G.\ {\bf 33}, 1381 (2001).

\bibitem{buchert:onaverage}
T. Buchert, in: {\em 9th JGRG Meeting, 
Hiroshima 1999}, Y. Eriguchi et al. (eds.), pp. 306--321 (2000).

\bibitem{buchertcarforaPRL}
T. Buchert and M. Carfora, Phys.\ Rev.\ Lett. {\bf 90}, 31101-1-4 (2003).

\bibitem{buchert:average}
T. Buchert and J. Ehlers, Astron.\ Astrophys. {\bf 320}, 1 (1997).

\bibitem{cover:entropy}
T.~M. Cover and J.~A. Thomas, {\em Elements of Information Theory} 
Wiley, N.Y. (1991).

\bibitem{hosoya:infoentropy}
A. Hosoya, T. Buchert, and M. Morita, in prep. (2004).


\bibitem{kullback}
S. Kullback, {\em Information Theory and Statistics}, Wiley, N.Y. (1959).

\bibitem{kullback_leibler}
S. Kullback,  R.~A. Leibler, {\em On information and sufficiency},
Ann.\ Math.\ Statistics  {\bf 22} (1951), pp 79--86.

\bibitem{WMAP}
D.~N. Spergel et al. Ap.J. Suppl. {\bf 148}, 175 (2003).


\end{thebibliography}
\end{document}